# How Physics Students Develop Disciplinary Computational Literacy

Tor Ole B. Odden[1*] and Benjamin Zwickl[1,2]

*[1] Center for Computing in Science Education, Department of Physics, University of Oslo, 0316 Oslo, Norway*
*[2] School of Physics and Astronomy, Rochester Institute of Technology, 84 Lomb Memorial Drive, Rochester, NY, 14607*
*[t.o.b.odden@fys.uio.no](t.o.b.odden@fys.uio.no) (corresponding author)

**Abstract**

Computation has revolutionized science and is gradually making its way into science teaching and learning. However, we currently lack theoretical frameworks to make sense of how students learn to use computation as a disciplinary tool. In this study, we propose *disciplinary computational literacy* as a productive theoretical lens on this subject. This theoretical perspective views computation as a new type of literacy consisting of material, cognitive, and social elements. We argue that these elements will necessarily vary by discipline and use case studies of two pairs of students writing computational essays in an intermediate physics course to examine how disciplinary computational literacy looks and is built at an undergraduate level. Through these case studies we see the pairs leveraging the different elements of their computational literacy to both engage with and produce computational literature in their discipline, while also engaging in a process of epistemic negotiation between their interests, the course goals, available tools, and their basis of computational literacy. These cases show how computational literacy can highlight the ways in which scientific computing helps students build disciplinary understanding and also shows that computational essays, as a genre of computational literature, are a useful epistemic form for developing computational literacy.

**Introduction: Computation has changed the ways we do and think about science**

Over the last century, computation has revolutionized what it means to do science (Denning & Tedre, 2019). Computational simulations are now widely considered a 3[rd] pillar of science, on equal footing with theory and experiment (Skuse, 2019). Computation has allowed researchers to solve large-scale, grand-challenge problems, has made it possible to visualize previously-intractable datasets and phenomena, and has paved the way for several recent Nobel prizes ("A Computational Perspective on the Nobel Prize," 2022). Machine learning methods are now allowing scientists to create programs that can parse unimaginably large datasets and make arbitrarily accurate predictions, as well as to investigate systems for which there might be no analytic solution. Computation has spawned new, hybrid fields of research like bioinformatics, computational physics, and social data science. And, across all fields, computation has revolutionized data sharing and led to dramatic changes in the ways scientists collaborate and communicate.

In light of these changes, it seems clear that future scientists will need to be trained in computational methods. Science students will need to learn basic coding practices in order to access these new ways of thinking and working (Denning & Tedre, 2019) and to use this technology to its fullest potential. Students also need to learn standard methods for tackling problems within their discipline: for example, solving differential equations (physics), agent-based modeling (epidemiology), and text analysis methods (social science). And, students will need to learn how to collaborate and communicate using these tools.

In response to this need, computation is gradually making its way into science education, both at the pre-college and collegiate level. Pre-college, many countries have added computational thinking, informatics/computer science, or programming to their educational standards (Bocconi et al., 2022; NGSS Lead States, 2013), and researchers have spent decades developing various innovative pedagogical approaches to integrating computation into science teaching (e.g., Hardy et al., 2020; Hutchins et al., 2020; Pei et al., 2018; Sengupta et al., 2013; Weller et al., 2022; Wilensky & Papert, 2010). At the University level, computation is being taken up both as a professional practice in existing science disciplines (e.g., Caballero & Merner, 2018; Haraldsrud & Odden, 2023; Magana et al., 2016; Odden & Caballero, 2023; Vieira et al., 2021) and taught in interdisciplinary data science degrees and courses (e.g., Silvia et al., 2019)

Thus, there is a critical need to understand how science students and educators learn to use computation as a tool within their disciplines. Existing theoretical frameworks, like constructionism (Kafai, 2006; Phillips et al., 2023), computational thinking (Denning & Tedre, 2019; Grover & Pea, 2013; Weintrop et al., 2016), and computational modeling (Buffler et al., 2008; Hestenes, 1987) provide useful starting points. However, these frameworks have historically focused on specific communities or tasks (e.g., maker spaces, problem solving, and physics modeling) rather than the holistic phenomenon of how computation is learned and used in disciplinary modes of knowledge construction.

With this in mind, the current study is focused on the development of a theory of disciplinary computational learning, situated within the development of a tool to help learn and assess computational knowledge construction. The theory is *disciplinary computational literacy*, an adaptation of the more general theory of computational literacy described by diSessa (diSessa, 2000) and Berland (Berland, 2016). The tool is a computational essay (Odden et al., 2019, 2022; Odden & Burk, 2020; Odden & Malthe-Sørenssen, 2021), a recently-developed multimodal communication genre that is commonly used in professional scientific computing practice, but has not yet seen widespread adoption within formal learning environments.

## Theoretical Perspectives

### Computational Literacy

Computation has long been seen as an essential tool for knowledge construction and a transformative agent for teaching and learning. It has often been compared with reading, writing, and mathematics as a core skill for knowledge acquisition (Forscyth, 1968; Kay 1984; Papert, 1981; Guzdial 2019). In other words, over the last decades computation has acquired the status of a literacy, in the sense of a general skillset necessary to life in the 21st century, similar to basic literacy in reading and writing or mathematics. This shift has led to its inclusion in various national curricula (Bocconi et al., 2022; NGSS Lead States, 2013).

However, as diSessa (2000) argues, there exist deep structural similarities that unite the modalities of the written word, mathematics, and computation and make them such powerful tools for knowledge construction. Each has a specific representational form: print literacy uses letters, words, and sentences; mathematics uses mathematical notation; and computation is based on computer code. These representational forms follow specific rules—grammar, mathematical logic, and syntax respectively—and each can be used to communicate ideas and accomplish intellectual tasks. When combined with disciplinary knowledge, these modalities allow people to solve previously-intractable problems: for example, predicting the motion of objects using calculus, storing and distributing religious knowledge using the printing press, or using machine learning models to analyze the human genome (diSessa 2018).

diSessa (2000) further argues that there are three essential elements, or pillars, of computational literacy that one must acquire to become computationally literate: material, cognitive, and social.

The *material pillar* is named for the idea that computational literacy, like other literacies, is based on a "material intelligence": the embedding of cognition into external signs, representational systems, and tools. This notion, which fits under the broader theoretical framework of distributed cognition (Cole & Engeström, 1993; Hollan et al., 2000; Hutchins, 1995; Pea, 1997), conceptualizes the representational system of computer code as a tool that can extend human thought and capabilities. However, computer code has certain affordances that differentiate it from other tools often discussed in the situated cognition literature, like calculators or cockpit instruments. Specifically, it is a system for *representation*—that is, a written language. It encodes certain types of information—instructions to a computer—and, once written, the lines, chunks, or scripts of code can be manipulated in various ways. At the same time, code is *executable*: if correctly written and run on a computer with appropriate software, it can send, receive, or modify data. Thus, as diSessa states, computational literacy is technologically-dependent; it relies on having a computer to do the execution.

Furthermore, as diSessa notes, code is a relatively recent human invention. And, although there are certainly carry-overs from its historical development (for example, the fact that most code is English-based or the genres of functional vs. object-oriented programming), it has been designed for a specific purpose—at its most basic level, performing logical operations and numerical calculations. Over time, and with the development of other related technologies like displays, internet browsers, and cloud computing, those calculations have allowed for manifold other applications, each with their own specific programming language (for example, HTML for web development, SQL for database management, Mathematica for mathematical modeling, etc.). Thus, the rapid design and proliferation of different varieties of code differentiates computational literacy from other material literacies that evolved over much longer timescales (like written language).

Becoming literate in the material aspects of computation involves learning the tools and techniques necessary to interact with this representational system, independent of domain or application. This includes the basic logic and syntax of specific programming languages; how to write, execute, and document code; common algorithms and data structures; use of developer environments; and so on. Thus, acquiring material computational literacy is comparable to learning foundational mathematical or writing skills.

The *cognitive pillar* consists of the modes of thought and interaction that are enabled by the material system of code—that is, the ways in which computational tools allow humans to solve tasks, express ourselves, and interact with the world. As diSessa (2018) argues, the representational systems that underlie literacies allow people to structure ideas, solve problems, and perform tasks that would otherwise be very difficult. For example, the formalism of calculus allows people to use rates of change as a variable to analyze constantly changing quantities, thereby describing many complex scientific problems and phenomena. The written word allows for argumentation and communication in formal and informal settings. And, similarly, computation allows people to perform calculations, analyses, and simulations that would be intractable or practically impossible to do by hand.

Theoretically, when focused narrowly on using computation for problem-solving, this pillar overlaps significantly with what has become known as computational thinking (Berland 2016; Wing, 2006), defined by Wing as "formulating problems and their solutions so that the solutions are represented in a form that can be effectively carried out by an information-processing agent" (Wing, 2010). However, as Kafai et al. argue, computation is also used to produce shareable digital artifacts for personal expression, and to critically engage with socio-political issues (Kafai et al. 2019). Thus, the cognitive pillar encompasses computational thinking but also covers the many other creative and analytical applications of computation.

Becoming literate in the cognitive pillar involves learning how to use the material aspects of computation to accomplish tasks, solve problems, and interact with the world. Importantly, this is different than using pre-made digital artifacts, in the same way that performing a statistical analysis is different than looking up statistics in a database, or writing a letter to an editor is different than reading a newspaper article. Using pre-made digital artifacts (e.g., spreadsheet software or web-design tools) is certainly an important skill, sometimes referred to as digital literacy (Blikstein, 2018), but the cognitive pillar of computation literacy requires greater facility with the underlying code and thereby opens for more flexible use of computation. Many people in today's society have digital literacy; fewer have computational literacy.

The *social pillar* consists of the ways in which we communicate with and about computation and code. Computational development has always been a fundamentally social endeavor. Hardware and software are developed by teams and companies, and since its inception software has been designed to be fundamentally shareable, for example via punch cards, magazines, cassette tapes, radio broadcasts, code repositories, or open-source software libraries. Most major software-based projects (both in industry and research) require large collaborating teams. Thus, as diSessa, and later Vee (2017), argue, computational literacy will always have an inherent social dimension. As Vee summarizes:

> "All code is embedded in human social contexts. Few programmers work alone, communicating only with their computer, and many of them program in agile environments where their work is subject to regular group review or to the instant collaborative work of pair programming. Even those few programmers who do work alone use programming languages that have been shaped by human history of programming and devices embedded in social histories. Programming—like writing—is a complex, social, expressive activity within a symbolic and technological system" (p.137)

This pillar operates on several different scales, from the macro-scale ways in which computation facilitates social interaction and communication to the specific ways in which people document, share, and collaborate on code. At an individual level, becoming literate in this pillar of computation therefore requires one to learn how to discuss, communicate, and share one's computational tools, problems, and results with others, and to collaborate on computational projects and analyses. However, these skills do not follow automatically with material and cognitive computational literacy. As Berland (2016) warns, "articulating meaningfully about programming is difficult: it requires overlapping teaching skills, speaking skills, and technical skills. Having some of those skills does not imply that you have all of them: many people who have significant programming ability often lack the ability to articulate the meaning and process of creating computational artifacts." (Berland, 2016, p. 197).

**Disciplinary aspects of computational literacy**

Learning scientists have long studied how physical and intellectual tools are used in different communities (J. Lave, 1991). For example, the mathematics used by professional mathematicians looks very different from that used by nurses (Hoyles et al., 2001), tailors (Reed & Lave, 1979), grocery shoppers (J. Lave et al., 1984), or Brazilian children selling candy (Saxe, 1988). This literature would suggest that it is naive to look for a single, unitary computational literacy—rather, there are likely to be many different "multiliteracies" spread across the different communities that use computation (Vogel et al., 2020).

Theoretically, we would therefore expect disciplinary differences across all three pillars of computational literacy. For instance, material aspects clearly differ by discipline, in that different communities use a diverse set of programming languages (Python, SQL, HTML, etc.) depending on their application, and have different ways of programming (developer environments, software libraries, program structures, and practices like pair-programming) within these languages. Thus, for example, scientists may use Python for modeling and data analysis, while software developers may use more efficient languages like C++ for commercial applications.

Cognitive aspects also differ by discipline, since different sectors have different goals, needs, and problems to solve: tech workers have different goals than scientists, journalists, or financial analysts, and what counts as computationally "knowledgeable" or "skilled" thus differs by discipline and application. Individual algorithms or numerical techniques can even serve different roles across disciplines: for example, numerical integration is a technique used to solve equations of motion in physics, rate equations in chemistry, population dynamics in biology, and financial algorithms in finance.

Social aspects also differ by discipline and application. Some sectors, like software development and large scientific endeavors, have a necessary focus on code sharing, distribution, and version control, since they require many people to collaborate on a single large codebase. Being literate in these domains requires significant skill in collaboration and communication. Other disciplines or applications, which rely on smaller teams and codebases, focus more on documentation or presentation of code results to peers.

These differences will necessarily be reflected in educational settings. For example, the specific elements required to educate computationally literate engineers (Magana et al., 2016), will look different from those required to educate physicists (Odden et al., 2019; Odden & Caballero, 2023), or biologists (Farrell & Carey, 2018). Understanding these disciplinary differences can help educators decide which computational tools and exercises are most appropriate to introduce into courses or degree programs.

**Computational Literature and Computational Essays**

Inherent to the theory of computation as a literacy is the notion that people who have acquired this literacy will engage with and produce a kind of computational "literature". And, indeed, diSessa (2018) argues exactly this:

> "Literacies shift the basic intellectual structure of domains of knowledge along with learning trajectories and societal participation structures—who gets to do what.

Concomitantly, *a literacy needs a literature*. One needs to transcend a representational system by itself, and get to civilizations' expanse of deep and powerful ideas." (diSessa, 2018, p. 7; emphasis added).

Prior work has identified several examples of students engaging with this new and developing literature. For example, diSessa (2018) describes a case study of a class of 6th-grade students using computation to explore basic concepts of calculus and Galilean motion using computational tools. He argues that this computational representation of motion is an example of students engaging with computational literature. Wilensky and Papert (2010) describe similar cases in which students use agent-based modeling to understand thermodynamic behaviors of gasses and crystal growth. Hutchins et al. (2020) describe how high school students used block-based programming to model the physics of projectile motion, such as a flying drone dropping packages. And Orban and Teeling-Smith (2020) describe how computational modeling activities can help introductory physics students analyze the physics of collisions. All of these cases exemplify how students can use computation to get at different "deep and powerful ideas."

When we look at professional practice, we see additional examples of these emerging literatures. For instance, one relatively recent development within computational professions and communities is the use of a new structure for programming and communication known as a *computational notebook*. A notebook is a tool that can be used to both execute code and situate it within a multimodal digital environment where one can write text and equations and embed images and videos. In certain cases, like open-source Jupyter notebooks (Kluyver et al., 2016), the code can also be broken up into different sections, or "cells" which can be run independently of one another, allowing writers to test and tinker with variations on their code while avoiding having to re-execute an entire script with each modification.

Although computational notebooks have been available under license for many years (Wolfram Research, Inc., 2019), recently free, open-source versions have emerged in most of the major programming languages. This has allowed scientists and data professionals to begin to communicate within this new medium, both for sharing of new ideas and communication of established projects and results (Rule et al., 2018). For example, computational physicists use notebooks to share data analysis "tutorials" like the gravity wave detection tutorial (LIGO Scientific Collaboration, 2019). They also use them to communicate results within research groups, and commonly publish them as addenda to scientific papers. The data science community has also widely adopted notebooks for communicating ideas, results, and methods. For example, the website Towards Data Science hosts thousands of short data science tutorials and essays that are written in and around notebooks. Thus, notebooks are forming the basis for a new kind of literature which natively uses text and computer code to communicate ideas and analyses.

As with any literature, the emerging computational literature encompasses many different genres of communication. One of these genres, which has been previously investigated in the educational research literature, is known as a "computational essay" (Odden et al., 2019, 2022; Odden & Burk, 2020; Odden & Malthe-Sørenssen, 2021; Somers, 2018; Wolfram, 2017). Here, we define a computational essay as a document that mixes text and code in order to describe an investigation, explain an idea, or tell a story. Computational essays are different from standard essays in that they include live, executable, modifiable code as part of their structure. This makes them ideal vehicles for communicating the methods and results of a computational investigation, since any reader with the requisite technology (a computer with an installation of the programming language) can see, execute, and verify that the results work as advertised. Computational essays provide an excellent modality for communicating computational scientific work, augmenting or even potentially replacing the standard scientific paper or report (Somers, 2018). They have even seen some adoption within the popular science realm—for example, Physicist Rhett Allain's writings at the magazine Wired frequently intersperse small, interactive Python applets to illustrate an explanation (Allain, 2017).

Computational essays, as a medium, also exemplify all three pillars of computational literacy. They incorporate code (material pillar) as part of their narrative structure in order to perform a disciplinary analysis or present an argument (cognitive pillar). The argument or analysis of a computational essay is frequently illustrated by the output of the code, such as plots or graphs, but the code itself must also be described, commented, documented, or otherwise explained in order to allow readers to understand that argument (social pillar). Computational essays therefore serve both a practical function, allowing one to execute code and run an analysis (material/cognitive pillar) and a communicative function, helping others to understand that analysis (social pillar).

## Research Questions

Because computation has only recently seen widespread adoption across different sectors and fields, we do not yet understand how computational literacy varies across different communities, domains, and disciplines. Nor do we know how people acquire these different computational literacies—how they interact, how they are similar or different, and how elements of literacy in one domain do or do not transfer to others. We are thus interested in the question of how students develop *disciplinary computational literacy:* the set of material, cognitive, and social elements needed to be able to use programming within a specific discipline.

Furthermore, computational essays have so far seen little use within the educational sphere. At the same time, it seems clear that computational essays, on their face, represent an increasingly important genre within the emerging disciplinary computational literatures.

Based on these observations, our research questions are as follows:

1. How does disciplinary computational literacy development look at an undergraduate level?
2. What role can computational essays play in supporting students' development of disciplinary computational literacy?

## Methods

### Educational Context and Design

The context for the present study is the physics department of the University of Oslo, Norway. The University of Oslo has a long history of integrating computation into their math and science courses (Odden & Malthe-Sørenssen, 2021). Nearly all math and science majors take a programming course during their first year of university studies, and computation is deeply integrated into several of the mathematics and natural science curricula. Physics, in particular, has a thorough integration of computation into teaching, with physics students taking both a scientific programming course and a numerical methods course (which focuses on teaching basic techniques for using computation to solve scientific problems) during their first semester. Subsequent courses build on this foundation, introducing more complex computational and data analysis methods. Thus, the University of Oslo provides a rich context for this study, in which students are actively building their computational literacy as part of their education.

Specifically, in this study we focus on students in a 3rd-semester physics course covering electricity and magnetism, which frequently included computational methods as part of the instruction and assessment. In this course, all students were required to complete an open-ended project during the second half of the semester during which they conducted a small, independent computational research project. In the course of the project, students (working either individually or in pairs) defined a question they would like to investigate, used a computational simulation to investigate it, and then wrote a Jupyter notebook-based computational essay describing their question, analysis, and results. They then orally presented their computational essays to a group of their peers, giving presentations of 8 or 13 minutes in length (for individuals or pairs, respectively) in a mock research-group meeting format, including 2 minutes for questions from the audience or TA.

In order to support students in this process, instructors provided students with several forms of scaffolding and support, which included the following:

*1. Guidelines for Professional Practice*

Students were provided with several types of guidelines on what they would be expected to produce in this project. First, they were given a written assignment description explaining the motivation behind the project and framing it as an opportunity for students to try their hand at an authentic, research-like scientific activity. This description also laid out the expectation for the procedure they would likely follow in developing their projects, the computational essay they would turn in, and the expected time commitment.

Additionally, students were given a copy of the rubric used in grading their computational essays, which featured 5 primary categories, each on a scale of 0-4:

1. **Investigation question:** The essay has an investigation question, it is physically meaningful, and it requires significant additions to an example simulation to answer.
2. **Coding:** The code runs without error and there are significant additions to any example simulation code used in the analysis (i.e., addition of new analysis methods, physical entities, or physics principles added to the simulation).
3. **Physics in the simulation:** Electricity and magentism-based physics principles have been used to augment any example simulation used, and it is clear how they were derived and applied in the code.
4. **Conclusions:** The essay has a conclusion which describes the results, interprets their meaning, uses them to answer the original question, and justifies their reasonability.
5. **Written report:** The written computational essay clearly explains the steps of the investigation, and includes at least 1 picture or diagram beyond the pictures/diagrams given in the original simulation.

In the project description, it was specified that the students would need to score at least 70% (14/20 possible points) on this rubric in order to pass.

*2. Supports and Examples of the Computational Literature*

In order to help students learn the genre of computational literature they were expected to produce, students were provided with three additional forms for scaffolding. First, examples of complete computational essays, written by faculty and previous students in the course. These example essays were explicitly designed to showcase best practices of computational essay writing: for example, a strong investigative narrative, judicious use of images and diagrams, iterative development of a computational model, use of functions and commenting for code cleanliness and documentation, citing sources, making reasonability checks, and explicit discussion of model limitations.[1]

Second, students were given access to a number of pre-made example simulations that were intended to act as "seeds" for the students to build out into fully-fledged computational essays. These simulations were stripped-down models of physics phenomena, written in Jupyter notebooks, that ran without errors and provided an overview of the basic theory behind the phenomena, but did not illustrate any especially interesting results. Example simulations included computational models of storm clouds, lightning (in 2D), a particle accelerator (cyclotron), a railgun, and a magnetic bottle, firing neurons, polar molecules in a liquid, and an electric dipole antenna. Each of these example simulations also included some suggestions for questions the students could pursue, such as "what are the effects of special relativity on the cyclotron simulation?" or "A reversed magnetic bottle is known as a 'bionic cusp.' How does a particle's behavior in a bionic cusp differ from that in a magnetic bottle?" However, students were also encouraged to build off of computational assignments they had done for homework, and/or write their own simulations from scratch if they felt inspired to do so.

The intention behind these simulations was three-fold: First, they provided an additional layer of scientific authenticity, since professional scientists usually build their analyses off of existing code or simulations and seldom write their programs from scratch. Second, they helped reduce the challenge of finding a question to investigate, which students often cited as one of the most difficult parts of the project (Odden & Malthe-Sørenssen, 2021). Third, by situating these example simulations in Jupyter notebooks, students were implicitly introduced to notebook use and encouraged to use them in developing their code and models.

The final scaffolding consisted of 2-hour drop-in sessions held by course TAs during the periods when students were working on their computational essays. Here, students could ask questions about theory, code, or logistics related to the project. TAs also provided feedback on student project ideas and simulation results.

**Data Collection and Analysis**

In order to answer our research questions, it was essential to understand the processes students used to write computational essays—that is, how they leveraged their pre-existing computational literacy and the provided supports to approach the project, and how they built their computational literacy in the course of writing the essay. For this reason, we chose a research design in which we gathered *process data*: that is, we followed multiple groups of students through the process of brainstorming, carrying out, and presenting their projects.

During the Fall semester of 2020, students enrolled in the course were contacted and recruited to take part in a series of interviews about their experiences with writing computational essays.[2] Seven groups responded and consented to take part in the project, and all participated in a series of interviews, with the first taking place before the students had begun work on their project, 2-3 interviews while they were actively working on it, and one interview after they had finished. Interviews were conducted over Zoom, by the first author or a research assistant, in either Norwegian or English. The interview protocol varied for each interview in the sequence, with the first interview focusing on background and initial ideas, the middle interviews focusing on progress and challenges during the project, and the final interview focusing on reflections on the essay-writing process. Additionally, students provided the researcher team with copies of their works-in-progress and (eventually) final computational essays after each interview. Interviews were subsequently transcribed, and all names and identifying information were replaced with pseudonyms and/or deleted. Interviewers also compiled research notes and analytic memos immediately after each interview focused on themes related to computational literacy and computational essay development.

---

[1] The essays are available at the University of Oslo digital computational essay showroom (Center for Computing in Science Education, 2019).

    Due to the exploratory nature of this study, our analysis used a narrative case study methodology paired with conjecture mapping (Sandoval, 2014), which proceeded as follows: first, the first author reviewed all transcripts and analytic notes to identify large-scale themes related to disciplinary computational literacy across the seven cases. Next, he chose two specific cases to focus on, which illustrated rich-but-contrasting examples of disciplinary computational literacy use and development. Both sets of interviews had taken place entirely in English. Transcripts and artifacts for these two cases were then shared with the second author, and both authors independently analyzed and inductively coded the transcripts and final computational essays in Dedoose (a software platform for collaborative qualitative analysis) with a focus on the three pillars of computational literacy, important aspects of students computational essay-writing process, use of supporting materials, and stated outcomes. Codes were then compared and discussed to sift out key themes.

    Based on these transcripts, analytic notes, and discussions, the first author then wrote narrative descriptions of each of the two cases, and these narratives were again discussed in light of the theoretical framework for disciplinary computational literacy—material, cognitive, and social elements—as well as the pairs' engagement with the computational literature. These narratives were triangulated with the analysis of the final computational essays.

    Simultaneously, the second author constructed graphical representations of the students' essay writing processes based on the conjecture-mapping approach described by Sandoval (2014). This approach examines the effects of educational designs by formulating and analyzing a high-level conjecture about the design. In the present study, this conjecture was as follows: *computational essays can support student development of material, cognitive, and social computational literacy in physics.* The conjecture map provides a way to evaluate this conjecture by capturing the relationships between three different elements of the learning process that takes place within said educational design: the *embodiment,* which consists of the ways in which the desired form of learning is codified in the design decisions and provided supports of the learning environment; *mediating processes*, which are the ways learners use these embodied elements along with their prior knowledge to accomplish tasks; and *outcomes*, which are the forms of learning that come out of the educational design. Due to the focus in this study on how students

---

[2] Although this study took place during the COVID-19 pandemic, teaching was conducted in a hybrid fashion, with a mix of digital lectures and in-person problem-solving sessions. The project was approved by the Norwegian Data Privacy authority (NSD), project #516136.

build computational literacy, in addition to the category of embodiment we also added a category of *prior experience*, meant to capture students' pre-existing elements of computational literacy going into the project.

Elaborate conjecture maps for each group were initially constructed based on all coded elements from the transcripts and essays. However, once initially drafted, these maps were discussed, triangulated with elements from the narrative case studies, and condensed. Simultaneously, the narratives were adjusted to incorporate elements from the conjecture maps that had been missed or downplayed during initial drafts. In this way, the two representations were iteratively brought into alignment with each other. Finally, the narratives were merged with each other to highlight similarities and differences between the two groups' approaches, and analytic reflections were added to the narratives to unpack the students' development of disciplinary computational literacy at key places.

## Results

### Description of Cases

#### Norah and Natasha

Norah and Natasha were 3rd-semester students in the University of Oslo physics and astronomy program. Both had learned the majority of their programming through their university studies, during which they had taken the standard coursework for their degree program: an object-oriented programming course in Python and courses in numerical modeling and mechanics (with an emphasis on computational modeling). Additionally, on top of the electricity and magnetism course that featured the computational essay design, they also were co-enrolled in an astronomy course with a heavy computational component. They thus came in to the project with a substantial base of disciplinary computational literacy in physics: material skills in Python programming, including familiarity with basic tools like variables, loops, functions, and objects; cognitive skills in using computation for physics simulation, data analysis, and problem-solving, for instance numerically solving equations of motion for projectiles; modeling specific phenomena from mechanics, electromagnetism, and astrophysics, for example air drag on moving objects, bouncing gas molecules, and electric potentials; plotting and visualizing results; interpreting the results of models; and some experience with notebook use and collaboration with other students on computational physics problems.

Despite this background, neither Norah nor Natasha reported an especially strong interest in programming during their initial interview, and they found the level of analysis in their coursework challenging:

> **Norah:** I *know that I'm not a big fan of coding in general, or I never was. That's why I'm a bit reluctant when it comes to working with codes.*
>
> **Natasha:** *And, I actually kind of feel the same way.*

However, they did feel that computation was a useful tool for building understanding in physics and astronomy, and they specifically appreciated the affordances of computation for visualizing physics phenomena and giving them easy ways to do rote calculations.

At the time of the first interview, the pair had started vaguely planning their project but had not decided on any specifics. They had a general interest in biophysics, inspired by a homework assignment in their electromagnetism course on the physics of nerve cells. They also had a strong desire to conduct a meaningful, realistic analysis which would give them some insight into real-world systems or phenomena:

> **Norah:** *And maybe, like, make it applicable to like real world problems. So, how it can be extended. So not just like in meaningless little physics and biology projects that you want to do, but something that can actually, like, bear some meaning. Or, I guess, more meaning into it.*
>
> **Natasha:** *Cure cancer.*

This interest in real-world problems would develop into a major theme of their project over the coming weeks, as the pair engaged with the computational literature they had been given and negotiated this interest with the constraints and requirements of the project.

### Linda and Amos

Linda and Amos were also physics and astronomy students, and had taken nearly identical coursework to Norah and Natasha—in fact, the four were friends and sometimes studied together. Linda and Amos thus came into the project with a similar base of computational literacy in physics, although Amos had a bit of prior programming experience, having done a year of engineering studies before enrolling in the physics program. However, the pair noted that it was their first time working on this type of open-ended project together.

In contrast to Norah and Natasha, both Linda and Amos stated that they enjoyed programming, and viewed it as a valuable skill for learning physics:

> **Linda:** *I would say for pretty much all math that we've done, it becomes a lot more understandable when you have to program it, because you have to really know what [it] does, what and why. And then you kind of understand everything you're doing more clearly, like explaining it to your computer*

At the time of the first interview, the pair had browsed through several of the example simulations but hadn't yet picked a topic to work on. They cited railguns and particle accelerators as interesting topics, due to their "coolness" factor.

> **Linda:** *We've been thinking about some topics. Like Railguns or particle accelerators. But we're not really sure which one to pick out of the models that we have been suggested. So yeah, I haven't really thought about it that much.*
>
> **Interviewer:** *Is there anything about, like, railguns or particle accelerators that stands out to you as being particularly interesting?*
>
> **Linda:** *For me I just think nuclear and particle physics is really cool. So it would be really cool to be able to implement that into electromagnetism.*

This "coolness" factor would also develop into a theme of their project over the coming weeks, leading to a different pattern of engagement and negotiation from Norah and Natasha.

**Defining research questions and planning analyses**

A week later (and a bit over a week before they were due to present their work) Norah and Natasha had identified the question that they wanted to investigate: how do intermolecular interactions between ions in nerve cells affect their movement? This choice of question was driven by one of the example simulations they had been given, a biophysics model which simulated ion motion in a nerve cell when the cell was firing. This model included a suggestion that students improve the simulation by adding interatomic potentials between the ions. This question appealed to the pair, since it had both a biophysics focus and would help develop the simplified model into something more realistic.

> **Natasha:** *We had to look at them [example simulation code] for a bit to understand exactly what was going on but it was really nice. And then we realized that there were the questions underneath that we could look at. And then we chose to look at the one where you add a potential. So instead of setting the force equal to zero and the particles come super close, we look at a potential. Because we realize that maybe it's a bit unrealistic and then we did some more research and then we realize that, yes, there are some things that happened between the particles and that we could look at the potentials that describe what happens. So we're thinking of doing that.*

In order to tackle this problem, the pair had spent significant time comprehending the code in the example simulation and trying to understand the underlying biophysical model, based on a combination of the simulation documentation and external resources like textbooks, Wikipedia, and YouTube videos. This required them to read up on several new topics that they had not previously encountered, like different types of interatomic potentials. However, despite a better understanding of the physics theory, the pair felt uncertain as to how they could implement these potentials in their model and evaluate and interpret their effects.

> **Natasha:** *It was a lot to understand. So we just started to like, look at the details in the code. And what does this mean? And what does it mean that he [the example simulation author], for example, when he looked at the... When the particles or the negative particles, charged particles, get close to the membrane... How... Yeah.*

Around the same time, Linda and Amos had abandoned particle accelerators and railguns in favor of the topic of lightning. This interest, they said, had come from an example simulation provided to the students which simulated a lightning strike in 2 dimensions on flat ground. One of the suggested questions asked what happens if one introduced a lightning rod or other conductor into the system. However, after a conversation with their teaching assistant about the scope of the project, they were planning on combining this simulation with another "cool" example they had seen in class which would allow them to simulate the lightning strike in 3D rather than 2D. Their specific question was formulated as follows:

> **Linda:** *we wanted to measure—we wanted to study what it would be like to get hit by lightning. So we wanted to measure the potential and the current of a man standing in a field with an umbrella and using the umbrella as a conductor. And also we wanted to find the likelihood of getting hit by said lightning. So we were measuring using this--or we*

> *were planning on using the lightning template that's on the website. [...] And Amos had a really good idea of plotting it and like doing it in 3D. So that could be pretty cool.*

In contrast to Norah and Natasha, Linda and Amos had spent little time reading up on the physics theory, since most of the concepts and numerical methods necessary for their investigation had already been covered in the course. Rather, they had spent significant time reading the example computational essays in the online showroom to understand the computational essay genre, and they planned to gain a better understanding of the physics underlying their model by exploring their simulation:

> **Linda:** *I'm interested to see how better my understanding of it will be when I start coding it and actually modifying the code myself, because I can kind of understand what the code is about now, but it becomes a lot clearer when you start having to move things around to get what you want. And yeah, I always find that coding helps me understand the topic a lot more. So I expect that to happen.*
>
> **Interviewer:** *Anything to add, Amos?*
>
> **Amos:** *Not really. I think, yeah. Same as Linda. Would be a little more clear once we actually start to modify the code and look into it. So far we've gotten, yeah, a little overview of what it means to write a computational essay. To begin with, I really had no idea what we had to do, so yeah. Yes. We learned something.*

In contrast to Norah and Natasha, Linda and Amos felt that it would be fairly straightforward to implement the planned changes to their chosen simulation:

> **Linda:** *I feel like from now on, it's going to be pretty straightforward. It's just going to be actually coding it and we've spent some time planning it. So it would be yeah, probably quite straightforward. And we're planning on working some more this weekend and I don't know, hopefully we'll be done by Monday or Tuesday.*

**Reflection: Both pairs begin by engaging with the computational literature**

In order to start on their projects, both pairs of students had to leverage their existing disciplinary computational literacy to engage with the computational literature they had been provided. That is, they had to leverage their material computational literacy to parse the code they had been given, and their cognitive computational literacy to interpret how the code simulated their respective physics phenomena. They further had to decide on an appropriate question that balanced the expectations of the course (learning electromagnetism) with their own interests and computational physics skills. Thus, it was their foundation of disciplinary computational literacy that allowed them to begin to engage with the computational literature and plan their analyses. Additionally, both groups took similar approaches to defining their investigation questions: exploring the example simulations and selecting one of the suggested questions. This was one of several possible approaches; other students chose to start their projects from scratch or adapt analyses from prior homework assignments.

However, at this point the pairs had also begun to diverge in their project approaches: Norah and Natasha spent much of their time trying to gain a deeper understanding of the phenomenon they were modeling, but struggled to determine how they would modify their chosen model to incorporate the factors they were learning about. In other words, they focused primarily on the cognitive aspects of their project, but these aspects were complicated by their struggles with the material aspects of the code. Linda and Amos seemed comfortable with both the material and cognitive aspects of their project, but had spent significantly more time reading example essays to comprehend the genre they would be writing in. Thus, they spent more time developing their social computational literacy.

We also note that in getting their projects underway, both pairs began a process of *epistemic negotiation*: that is, negotiation between their interests (as stated in the first interviews), the goal of the assignment as stated in the assignment description and rubric, the computational and theoretical tools at their disposal, and their existing basis of disciplinary computational literacy. All of these factors came into play during their efforts to identify an interesting-but-tractable question to investigate. Within this process of epistemic negotiation, further differences begin to emerge between the groups: Norah and Natasha chose to grapple with a complicated biophysical phenomenon, driven by their interest in both biophysics and realistic problems, which pushed them to spend their time learning new disciplinary content (reading up on potentials). Linda and Amos were more interested in addressing a "cool" topic (lightning strikes), and together the pair felt comfortable enough with their code and theory that they could directly implement it into their chosen example simulation.

**Developing and presenting their simulations**

A week later, both pairs were in the final stages of writing their computational essays and preparing to present their findings to their peers. Norah and Natasha had spent most of the intervening project-time implementing their modifications to their simulation, exploring the results in their model, and writing their report. By their own admission, most of the coding had been done in a 10-hour marathon session over the weekend. Because they were not especially familiar with notebooks, they chose to do their coding in a different developer environment and copy their code back into the notebook, which they planned to use for their computational essay and presentation. However, the pair reported that they had run into some trouble modifying the code in the way that they wanted, which necessitated some scaling back of their original plans:

> **Norah:** *We studied it very well and we thought we understood everything. And as we told you previously, it [the example simulation] was a brilliant code and we really liked it. But then the moment we wanted to implement something new, something apart from the force, which was quite easy to implement, we wanted to look at the electric field just because we thought that it would be nice to see if, whether the new forces would have an effect on that-*
>
> **Natasha:** *Or just something as easy as printing the forces.*
>
> **Norah:** *Yeah, exactly. Something as easy as printing the forces. We were not able to do that because the code was running in a loop, for some—not for some reason, we know it was running in a loop because it's doing the Euler-Cromer [numerical integration] for the number of time steps. And then for some reason the way that person structured the code, it was also printing out the force value as many times as the time step value, which was like 1,500 times. It's a lot. And we couldn't fix it because we couldn't code the*

> *function for the force separately, because he integrated in a way that was, everything was dependent on each other and you couldn't just isolate one bit and only work with it.*

Briefly summarized, the pair ran into trouble due to the structure of their chosen simulation, which was deliberately written in a computationally efficient way. The simulation was modular, leveraging several different computational functions to initialize the model, draw a visualization, simulate random motion of ions in the cell, and then add in electromagnetic interactions between the ions. Although this structure allowed the pair to more easily understand their model, it made it difficult for them to modify it since the model's functions depended on one another and the changes the pair had hoped to make required modification of several different functions.

In the end, they chose to make a relatively small change to the model code: copying, pasting, and renaming an existing function in the simulation, then adding several lines of code to simulate the different forces and commenting out those that weren't used, as shown in Figure 1. They worried that these changes were insufficient to receive a full score on the rubric category related to coding, but reassured themselves that as long as they received high scores across the other categories, they would still pass the project.

```python
    for i in range(num_charges):
        # Cycle through all charges
        mask = mask_indices != i # Masks/picks out charge number i
        r_vec = r[:,i].reshape(2,1) - r[:, mask]  # Distance vector
        # reshape allows us to subtract the current (x,y) from all other
 ↪positions
        r_norm = np.linalg.norm(r_vec, axis = 0) # length of vector
        lennard_jones = (((24*epsilon)/(r_norm)**2)*((2*(sigma/
 ↪r_norm)**12)-((sigma/r_norm)**6)))*r_vec #introducing the new force
        f_morse = (((2*a*De)/r_norm)*np.exp(-2*a*(r_norm-min_dist)))*(1 - np.
 ↪exp(a*(r_norm-min_dist))))*r_vec
        f = np.where( r_norm > min_dist, q[i]*q[mask]*r_vec/r_norm**3,0)
 ↪#original expression for force
        """Implementing new forces"""
        #f = np.where(r_norm > min_dist, q[i]*q[mask]*r_vec/
 ↪r_norm**3,lennart_jones) #testingsimulation with LJ-force
        #f = np.where(r_norm > min_dist, q[i]*q[mask]*r_vec/r_norm**3,f_crazy)
 ↪#testing simulation with Morse-force
```

**Figure 1:** Norah and Natasha's primary additions to their model code

On top of these difficulties with the code, Norah and Natasha had spent significant time trying to make sense of an anomalous result: when comparing the effects of two different types of interatomic potential models (Lennard Jones and Morse) on their simulation, they had seen almost no difference in their results.

> **Norah:** *When we plotted both of our, or when we ran the simulation for both, when the force was equal to zero and with Lennard Jones or the Morse potential, we saw that there was barely any difference or it wasn't substantial enough for us to discuss it and like say anything about it. But that's why we're like, okay, well something shady.*

This result remained puzzling until the next day, when (during their subsequent and final interview) they reported having recently achieved a conceptual breakthrough:

> **Norah:** *We couldn't sleep at night because I was thinking about the--those...*
>
> **Natasha:** *I called her this morning at the t-bane [metro train line].*
>
> **Norah:** *On the t-bane. "I know why it makes sense now, it's because..." And I was also sitting on the t-bane and I was like, "Yeah, yeah. Makes sense." And there were so many people around and I'm like, "Well, they must think that we're really weird."*
>
> **Interviewer:** *Could you tell me a little bit about the thing that clicked for you this morning?*
>
> **Norah:** *We had this Morse potential that we looked at. I don't know if you remember, and this did not make any difference when we implemented this new force from the Morse potential. And it didn't make any difference between this and the original code where he set the force to zero. And we didn't understand why, because we saw that when the distance get very small, the force will get a lot bigger. But we saw that the force from the Lennard Jones potential was--that made a difference.*

In essence, the pair had realized that the two potentials they were working with operated on different length-scales, with one only having an effect when simulated ions came very close together. This had initially led them to see very little change in their model behavior when adding in the new forces, but once they modified the threshold over which the new forces acted they were able to see significant differences in their simulation results.

This breakthrough was unpacked in detail in their final computational essay draft, where the pair spent significant space in their essay motivating, exploring, and describing the effects of the changes they had made to their model. Their essay featured a detailed theory section on the different potentials they were using including both mathematical and graphical descriptions; simulated seven different scenarios that explored the effects of their chosen potentials on the model behavior; and discussed these different scenarios, unpacking the differences observed across the simulated scenarios.

Linda and Amos had also completed most of their analysis by this point, and had nearly finished writing their computational essay despite some technical issues with code sharing and cloud servers.

> **Linda:** *We've done a lot since last time. We've written up everything, nearly. We're just a little bit finishing up some details and we have to finish the conclusion, but we've pretty much written up everything and finished the code. I think it's gone well.*

As they had predicted, it had been fairly straightforward for the pair to implement their model in code, but had some difficulty with explaining what they had done in the textual component of their essay.

> **Amos:** *Coding is fun, but the writing part is kind of hassle, getting everything down, making sure that it makes sense.*
>
> **Linda:** *Yeah. And explaining, like, in detail what you did, it's so much easier just to do it, and then we have to explain that, why you did it, and you kind of have to, yeah. You have to defend yourself a lot more, the reasons for doing things. Yeah. And making it sound nice and smooth in the way that you're explaining it is also such a hassle.*

The pair added that they had spent significantly more time on the writing of their essay than the computational modeling, but that they had consequently learned quite a bit about the physical mechanism that leads to lightning strikes:

> **Linda:** *I wrote a lot about the dielectric breakdown, where it creates—makes it so that the air could lead, like conduct the electricity so that the lightning could strike. It's not something I knew of from before, which was kind of cool.*

Their final essay draft reflected these choices: much of the narrative was dedicated to a detailed description of the numerical technique used to model lightning, with visualizations of the discretized grid technique they had used. The pair had made significantly more additions to the model code than Norah and Natasha's essay, but had spent much less space in their essay motivating the investigation question and interpreting results.

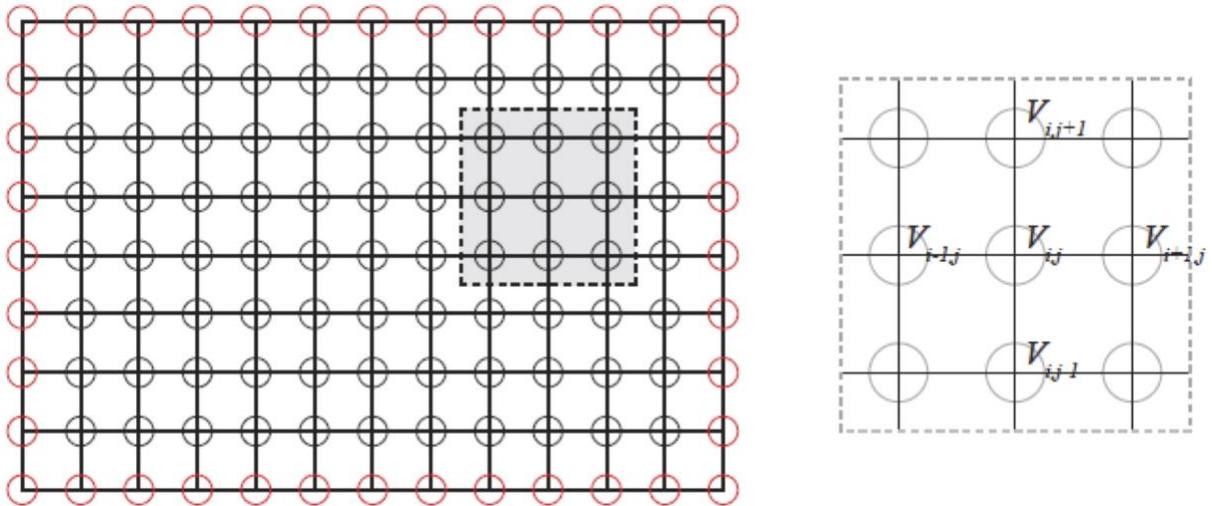

**Figure 2:** Linda and Amos' visualization of the discretized grid technique used to model the lightning strike

**Reflection: Production of disciplinary computational literature**
At this point in the project, the pairs had moved from engaging with computational literature written by others to producing their own pieces of computational literature. In this act of production, we can see their interests and disciplinary computational literacy reflected.
Norah and Natasha continued to focus on creating a realistic model, but this focus was constrained by their difficulties with the simulation code they were using—that is, the material computational aspects of their project. So, the pair spent significant time puzzling over the realism of their results (that is, an unexpected "null result") and reveled in this process of sensemaking, even calling each other the morning of their presentation to continue to discuss results. Thus, when it came time to write their computational essay the pair chose to focus on the cognitive and social aspects of the project—that is the motivation, interpretation, and communication of their model results—over the more material, code-focused aspects.

Linda and Amos had a significantly easier time modifying their chosen simulation, both because they were more comfortable with the material aspects of their code and because they were more familiar with the physics and numerical techniques underlying their model. But, they had a harder time interpreting and communicating the results of their investigation—that is, with the social and cognitive computational literacy aspects of the project. Consequently, they left this until the end, and focused much of their communication on the actual model itself, rather than the phenomenon or question under investigation.

In this way, although both groups succeeded at producing a piece of computational literature that blended material, cognitive, and social aspects of computational literacy, they did so in very different ways. Norah and Natasha's project is emblematic of how students can productively use computation to understand the world even with relatively small changes to a model or script. Linda and Amos' project shows how students can also productively focus on model development without going as deeply into model interpretation or communication.

**Student reflections on the process of writing computational essays**

The final interviews with each pair took place shortly after they had orally presented their computational essays to their peers. Norah and Natasha felt positive toward the experience as a whole: their presentation had gone well, they said, and they felt they had benefited from the experience of showcasing their computational essay. The TA running the session had even given them constructive feedback on presenting scientific results, and they found it interesting to see the work of another student who had pursued a similar topic.

When asked what they had learned from the process, Norah and Natasha cited a better grasp of the skills involved in physics modeling: creating, using, and evaluating models. At the same time, they felt that they had not gained significant understanding of any specific physics topics related to their coursework, despite the fact that they had spent substantial time learning about and modeling interatomic potentials (a topic clearly related to electricity and magnetism).

> **Interviewer:** *So thinking back on the process now, are there any aspects of electromagnetism that you feel like you understand better after you've gone through this process, or any aspects of physics as a whole?*
>
> **Norah:** *I think definitely looking at models mathematically and predicting what they do. I think we've gotten a much better intuition for that. Because when we began working with it in the beginning, we weren't sure about the results. We didn't really know-*
>
> **Natasha:** *And when we got the results, we were disappointed because we thought it'd be a very big difference. But it was not a difference we--yeah.*
>
> **Norah:** *Yeah, exactly. But then I think we also developed better skills regarding simulations in general, like looking at the code and really understanding what it's doing.*

> *Because that was crucial. In this case, you couldn't just predict it by looking at the equation. Because there are so many other things that are playing a role. So, I think that electromagnetism specifically, I don't think so.*
>
> **Natasha:** *I don't think so. But physics-*
>
> **Norah:** *Physics, in general physics? Yes. I think computational physics, yes.*

In their final interview, Linda and Amos also reported that their presentation had gone well, even though Linda had initially been somewhat nervous about presenting their results to peers. They had also received positive feedback from the presiding TA as well as affirmation from friends, including Norah and Natasha who were in the same presentation group:

> **Linda:** *It's a little bit stressful though, to have to present in front of people rather [than] just to hand in the essay. But at the same time it can be more rewarding. Our TA was very good at constructive criticism. He was quite nice. He was saying it in a nice way. I don't mean he was critical, but it was actually very helpful to get feedback on how to present and what to emphasize, which we wouldn't have gotten if we'd handed it in as an essay.*

At first, the pair reported that their final conclusions were fairly underwhelming:

> **Interviewer:** *So did you figure out anything new about what happens when somebody gets hit by lightning?*
>
> **Linda:** *Well, nothing new other than he--I mean, we found out he'll probably die. There's a good chance of that. That's pretty much it.*

However, when pressed, they added that they had, in fact, learned a significant amount about the physics of lightning strikes:

> **Interviewer:** *So, could you say a little bit more about that?*
>
> **Linda:** *We found the resistance of... Well we found out that the resistance of human skin when it's dry is about a hundred thousand ohm. And that ends up going away after--while the person gets hit by lightning. It decreases quite rapidly to a thousand ohm. Most of the resistance in the human body is in the skin itself, which is what I found. I also found out that women were more likely to survive lightning strikes. Although this didn't end up in the report, it was something to do with the area of women to men. So men had on average, a greater area and therefore lower resistance, I think.*

Like Norah and Natasha, Linda and Amos also reported a better understanding of computational physics presentation and simulation techniques:

> **Linda:** *What I learned the most probably was how to structure sort of like a scientific presentation. Because [Professor] told us that he wanted us to kind of imagine it being as* *if we present to our colleagues. And I think this is a useful thing to have done before and learn before you have to actually do it in front of your colleagues. I guess that was the most useful thing that I learned.*

**Amos:** *Yeah. Same thing. Maybe also the codes about how to calculate the potential. We had to figure out how to do it in three dimensions. So it helped a lot to understand the basics of the codes.*

**Summary of cases**

Figures 3 and 4 are compact graphical summaries of the trajectories of the two pairs through the project, based on the conjecture mapping technique (Sandoval, 2014). In representing these trajectories, however, we have made several modifications to that technique as described, in order to highlight the similarities and differences between these two groups. Specifically, (1) relevant prior experiences are included in the figures to explain how students with access to similar features of an embodied design and mediating processes can achieve different outcomes; (2) the features of the embodied design are highlighted in the textboxes of the mediating processes using bold font to highlight how the different embodied features were used; and (3) the mediating processes are not shown in parallel, but in a time-ordered sequence to better reflect the project narratives described above.

Figure 3 shows Natasha and Norah's conjecture map. From their prior experience, two features are particularly salient to their progression in the project: their self-perceived reluctance with coding, and their strong interest in biophysics and real-world problems. In the *Ideation* phase, they searched the provided example simulations for an applied problem, landing on an model that dealt with neurons and selecting one of the suggested questions from that simulation. The example simulation thus played an important role in their ideation phase, both defining the topic of their analysis (the biophysics of neurons) and the scope of their investigation question.

Next, the pair turned to their model. In the *Model comprehension* phase they spent significant time both comprehending the model code and the physics it was based on. Based on the understanding they gained, in the *Model refinement phase* the pair tried to modify the simulation but encountered difficulties due to the structure of their simulation's code, resulting in relatively minor modifications to their model. However, the pair were still able to find an interesting result to unpack and make sense of, which formed the core of their computational essay.

The final mediating process, *Essay writing*, gave the pair a space to share their knowledge and answer their research question. Because they were less familiar with Jupyter notebooks as a coding environment, they chose not to use them for code development only presentation. However, the presentation gave them both a chance to showcase their work for their peers and compare their project to another group who had pursued a similar question.

As a result of the project, the pair showed limited growth in their understanding of coding (material computational literacy); however, they engaged in sophisticated modeling practices and used their computational skills to gain insight into their chosen biophysics system (cognitive computational literacy). Their notebook also demonstrated an understanding of their audience and contained a thorough answer to their chosen investigation question (social computational literacy). Although the pair did not express any significant learning gains in their knowledge of electromagnetism, they did learn about both interatomic potentials and the more general process of computational modeling in physics (disciplinary knowledge).

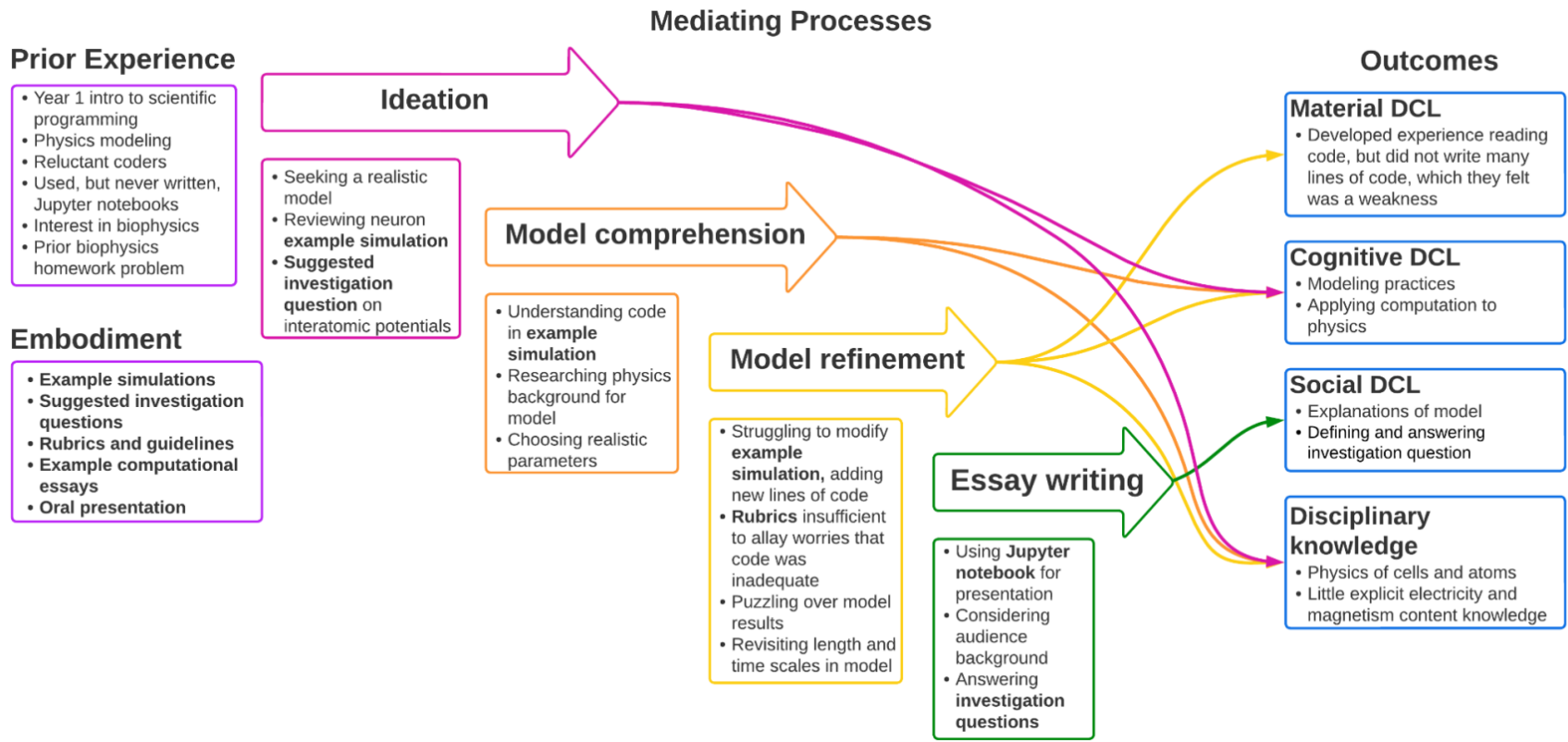

**Figure 3:** Natasha and Norah's conjecture map, evaluating the conjecture "Computational essays can support student development of material, cognitive, and social computational literacy in physics." Features of the embodiment are shown in bold text. The mediating processes are somewhat sequential and are arranged in timeline order. Outcomes, including disciplinary knowledge and material, cognitive, and social disciplinary computational literacy, are shown on the right.

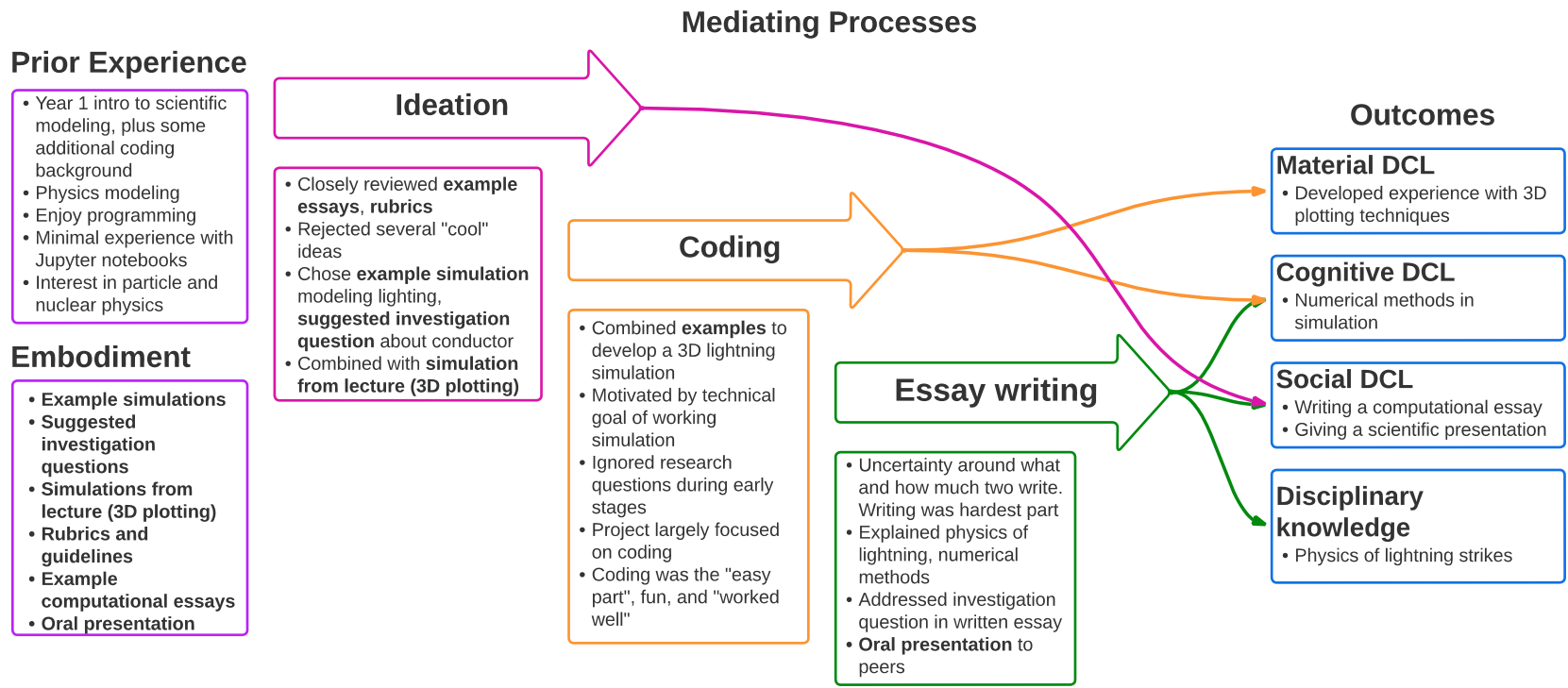

**Figure 4:** Linda and Amos' conjecture map, evaluating the conjecture "Computational essays can support student development of material, cognitive, and social computational literacy in physics." Features of the embodiment are shown in bold text. The mediating processes are somewhat sequential and are arranged in timeline order. Outcomes, including disciplinary knowledge and material, cognitive, and social disciplinary computational literacy, are shown on the right.

Figure 4 shows a similar conjecture map for Linda and Amos. As shown, the pair came into the project with a similar base of prior experience to Norah and Natasha, however the pair had a higher level of enjoyment and comfort with computation. For Linda and Amos, the *Ideation* phase of their project was less focused on answering an authentic question or gaining deep scientific insight, and more focused on getting an interesting or "cool" simulation to work. Although the pair initially selected a suggested question from the example simulation, it did not play a guiding role in their project. Instead, they focused on combining the example simulation of lightning with another simulation from class that gave them tools to do 3D plotting in order to make a model of a lightning strike in 3D.

For Linda and Amos, the *Coding* phase of their project was the self-described easy part, in which the pair implemented their planned changes to their example simulation and evaluated the results. Their conjecture map does not include a Model comprehension or Model building phase because most features of the physical model they presented in their final computational essay already existed in their chosen example simulation. In other words, the focus of their project was to improve their simulation at a technical level, which required both modification of their numerical method (bringing it from 2D to 3D) and their plotting and visualization methods; this did not, however, require them to spend significant effort comprehending or modifying the underlying physics of their model.

For Linda and Amos, the final mediating process, *Essay writing*, was the most difficult and time consuming. Although the pair felt comfortable coding and modifying their simulation, the computational essay required them to deeply understand and explain how the physics of lighting and numerical algorithms worked. It also required them to return to the question they had articulated early on, about what it would be like to be hit by lightning, and consider this question in light of their simulation. In this way, the mediating process of writing the essay pushed them out of their comfort zones and led to deeper learning.

As a result of their project, Linda and Amos gained a better understanding of 3D plotting techniques (material computational literacy), numerical methods (cognitive computational literacy), and some understanding of the physics of lightning (disciplinary knowledge). They also commented that this had been their first opportunity to both write a computational essay and give a presentation about their own research, including receiving and responding to feedback (social computational literacy).

**Reflection: Development of disciplinary computational literacy through production and presentation of computational literature**

If we examine their trajectories as a whole, both groups started with fairly comparable levels of pre-existing disciplinary computational literacy, even though there were some differences in interest and fluency. Both also approached the project in similar ways, starting by reviewing the example simulations provided to them, which required a combination of material and social computational literacy in physics. Both chose a suggested investigation question, eventually addressed their questions, produced comparable artifacts, and successfully presented their results to a group of their peers.

However, as both the narratives and conjecture maps show, there are clear differences in the approaches, trajectories, and outcomes for the two groups. Norah and Natasha had a strong interest in sensemaking and realistic modeling, but were less comfortable with coding. The pair spent most of their time learning physics theory to help them implement the changes they wanted to make in their model. For them, the essay provided a venue to communicate the breadth of their scientific work. If Norah and Natasha had been evaluated solely on their code, their project might have looked underwhelming, but when viewed in the context of what they learned about biophysical models, how they explored multiple simulated scenarios, and how they answered their investigation question, theirs was a substantial computational essay.

Linda and Amos were drawn to coding, but showed less interest in interpreting or making sense of their model results. The pair went directly to their code, implementing their model there. Compared to Norah and Natasha, they came to the project with higher interest in and comfort with coding and focused more of their efforts on new coding techniques (3D visualization) and understanding and their numerical methods. For them, the writing of the essay was the most difficult part of the project, yet that mediating process led them to a deeper understanding. At the same time, Linda and Amos also demonstrated growth in computational knowledge and skills, both in learning new visualization techniques and because they wanted to provide clear explanations for how their code worked.

Despite these differences, we see significant evidence that both pairs developed their disciplinary computational literacy. Both pairs had to leverage their pre-existing computational literacy in order to engage with the computational literature they had been given, determine a question to investigate, and decide on an approach to answering it using their chosen computational model. We also see how they had to stretch and build on this literacy to learn new ways of working with model code, and interpret and present results. The students themselves also saw this development and, when asked, explicitly commented on it; this, we would argue, is further evidence of computational literacy development.

An important aspect of the students' production of computational literature was their process of epistemic negotiation, in which both pairs tried to find a middle ground between their interests ("cure cancer"/"cool" simulations), their computational skills, the expectations of the course, and the models and tools available to them in order to define an investigation question that would be both interesting and tractable. This kind of negotiation is highly authentic to the discipline of physics—and, in fact, to all research disciplines. However, students are not often given opportunities to authentically engage in this kind of epistemic negotiation. Most exercises in standard physics and mathematics courses, even computational exercises, are fairly closed-ended. They require students to build and use specific analytic skills but do not give them the freedom to leverage their interests or choose their models and tools.

## Discussion
### Computational Essays as an Epistemic Form

Constructionists have long argued that the production and use of computational artifacts leads to computational fluency and computational thinking (Kafai, 2006). Thus, it is, from one perspective, not surprising that the students featured in this study built their disciplinary computational literacy through the production of computational literature.

At the same time, computational essays are quite different from other artifacts studied by constructionists, such as geometric shapes, robots, video games, or programmable textiles (Berland, 2016; Berland et al., 2013; Lui et al., 2020; Papert, 1980). We see four primary differences. First, computational essays explicitly embed disciplinary knowledge. That is, they are *epistemic* artifacts, which use code as a tool to produce and describe understanding within a discipline. To produce such epistemic artifacts the students in this study had to go through a process of epistemic negotiation, in which they directly engaged with the different pillars of their computational literacy. Thus, the epistemic nature of these computational essays was integral to the way in which the educational design supported students' development of disciplinary computational literacy.

Second, computational essays are based on the practices of a particular discipline—in this case physics—and they embed the particular approaches to inquiry, ways of knowing, and judgments about the quality of knowledge specific to that discipline. In the present context, physics knowledge often takes the form of computational models, and the results of such models are evaluated in light of physics theory and model limitations. These practices were clearly visible in the students' computational essay writing processes and the artifacts they produced.

Third, computational essays are synthetic—they require one to write code, but they also require prose, model descriptions, and visualizations. This again requires students to leverage all three pillars of computational literacy, and makes the process of writing a computational essay itself a synthetic activity. At the same time, different aspects can amplify or interfere with one another. For example, Norah and Natasha's development of their model was somewhat hampered by their ability to engage with their code, and Linda and Amos' interpretation of their results was driven by the need to present their project in their computational essay and oral presentation.

Finally, computational essays are flexible: although they draw from all three pillars of computational literacy, they also allow students to choose which aspects they would like to focus on. For Norah and Natasha, this meant focusing on biophysics topics, realistic modeling, and using computational simulations to make sense of the world, while Linda and Amos' focus was on code and model development, and opportunity to simulate something "cool."

Based on these observations, we argue that gaining computational literacy in a discipline—especially a knowledge-producing discipline like mathematics and natural sciences—requires producing pieces of computational literature, in the same way that becoming research literate involves engaging in the research process and producing a piece of research literature. As a corollary, this suggests that science educators who wish to deeply incorporate computation into their teaching would do well to use such epistemic forms as the basis for their instruction.

However, as shown in this study, producing disciplinary computational literature is not easy, and students will need support in order to successfully create such an epistemic form. In the case described above, the students leveraged many different forms of scaffolding: they read through example essays, assignment descriptions, and grading rubrics; built off of provided example simulations; pursued suggested research questions; and received help from teaching assistants. These supports themselves can be tuned to embed disciplinary ways of knowing and working. For example, professional researchers seldom write code from scratch or come up with research questions out of the blue; instead, they build off of other researchers' code and take up unanswered questions posed by the research community, just as the students did in this study. Professional researchers also write papers and reports in the same genre as the literature that they read; in the same way, the students in this study used the assignment guidelines, rubric, and example computational essays to understand the computational essay genre. Professional researchers present their results to peers in formal and informal settings, much like the mock research-group meetings within which the students presented their project results. In this way, these different elements can work together to support students in producing a piece of computational literature authentic to their discipline.

**The emergence of a new literacy**

Broadening our focus, we see this case study as emblematic of the new, emerging literacy described by diSessa (2000, 2018) and other authors like Berland (2016) and Vee (2017). This case shows that computational literacy involves much more than simply knowing how to code, or even how to engage in computational thinking—although both of these elements are essential, they need to be combined with disciplinary knowledge, communication, and the production of artifacts and computational literature to reach the potential for changing society that was predicted by Papert (1980), Wilensky and Papert (2010), diSessa (2018), and others. If our emphasis is simply on teaching kids to code (i.e., learning basic coding techniques like variables, loops, and functions) that will lead to a very different set of education practices and assessments than one would use to help students become computationally literate. Thus, we are arguing for an alternative educational approach, in which communities of professionals, educators, and learners work together to build up libraries of computational literature in their disciplines.

This case shows how students, at an intermediate level of disciplinary computational literacy, gained additional fluency through the production of a piece of computational literature. Andrea diSessa has stated that "a literacy needs a literature" (diSessa, 2018, p. 8) and this study speaks to this need. Further, although we have focused on computational literacy within the realm of physics, computation spans and transcends disciplinary boundaries. It is constantly being adapted to the needs and ways of thinking of different communities, in much the same way that mathematics has been integrated across all aspects of society. We see a great need to investigate how learners from other disciplines acquire computational literacy, what characterizes and differentiates those different literacies, and what kinds of computational literature can support this development.

With the recent development of generative AI tools like ChatGPT, these knowledge needs have become even more pressing. Such tools can enable users to easily generate and run code, which may help make computational literacy significantly more accessible to a wider variety of learners. At the same time, without a fundamental degree of disciplinary computational literacy, users will have difficulty evaluating the quality, output, or usefulness of AI-generated code. Such issues can become critical, even dangerous, in cases where users might be tempted to integrate AI-generated code into critical systems like engineering applications, logistical tools, or medical software. So, future studies of disciplinary computational literacy and computational literature will likely need to take into account the role of AI tools.

In summary, it is clear that disciplinary computing is not going away. Over the coming years and decades, as computational literacy becomes more widespread in the population and more and more disciplines incorporate computing into their educational systems and pathways, disciplinary variants of computational literacy will continue to emerge. This will likely accelerate both with the adoption of generative AI tools and as computation continues to be integrated into pre-college education systems worldwide (e.g., Bocconi et al., 2022; NGSS Lead States, 2013). Discipline-specific training programs will need to be developed to help with these emerging literacies, such as the ongoing effort to incorporate computation into physics education (Caballero et al., 2019; Caballero & Odden, 2024; Irving et al., 2017; Odden & Caballero, 2023; Phillips et al., 2023; Weller et al., 2022). We, as a community, have a great opportunity to inform and study this development.

**Conclusion**

It is a goal of the learning sciences to understand how learning happens. As has been argued many times, computation is an ideal tool for learning—or, as Papert put it, the computer is an "object to think with" (Papert, 1980). But, in order to deeply engage with this tool, people will need more than just programming skills or generic computational thinking strategies—they will need to understand how to fluently use computation to collaboratively solve meaningful problems in their lives and disciplines. It is up to the learning sciences community, as well as other fields like discipline-based education research, to deeply understand how this learning happens and how we can help people to become computationally literate in their disciplines.

## Acknowledgments

This work was funded by the Norwegian Agency for International Cooperation and Quality Enhancement in Higher Education (DIKU) which supports the Center for Computing in Science Education, and the U.S.-Norway Fulbright Foundation for Educational Exchange. We thank Anders Malthe-Sørenssen, Sebastian Gregorius Winther-Larsen, and Crina Damşa for their help and feedback with this project.